\documentclass[a4paper]{report}
\usepackage[utf8]{inputenc}
\usepackage[T1]{fontenc}
\usepackage{journal}
\usepackage{amsmath,amssymb,array}
\usepackage{booktabs}

\usepackage{longtable}

\newlength{\cslhangindent}
\newlength{\csllabelwidth}
\newlength{\cslentryspacingunit} % times entry-spacing
  {}%
  {\par}
 {% don't indent paragraphs
  \setlength{\parindent}{0pt}
  \ifodd #1
  \let\oldpar\par
  \def\par{\hangindent=\cslhangindent\oldpar}
  \fi
  \setlength{\parskip}{#2\cslentryspacingunit}
 }%
 {}
\usepackage{calc}

\usepackage{metalogo}
\begin{document}

\begin{article}

\title{Rendering LaTeX in R}

\author{by Paul Murrell}

\maketitle

\abstract{%
The xdvir package provides functions for rendering LaTeX fragments as labels, annotations, and data symbols in R plots. There are convenient high-level functions for rendering LaTeX fragments, including labels on ggplot2 plots, plus lower-level functions for more fine control over the separate authoring, typesetting, and rendering steps. There is support for making use of LaTeX packages, including TikZ graphics. The rendered LaTeX output is fully integrated with R graphics output in the sense that LaTeX output can be positioned and sized relative to R graphics output and vice versa.
}

\section{Introduction}\label{introduction}

Text labels, titles, and annotations are essential components of
any data visualization. Viewers focus a lot of their attention
on text \citep{massvis},
text is the most effective way to communicate some types of
information \citep{hearst},
and the message obtained from a data visualization can be
heavily influenced by the text on a plot \citep{kong}.

R provides relatively flexible tools for adding text labels to plots.
For example, in the \texttt{graphics} package, we can specify an
overall plot title and axis titles via the \texttt{main},
\texttt{xlab}, and \texttt{ylab} arguments to the \texttt{plot()} function
and we can add text at arbitrary locations on the plot with the \texttt{text()}
and \texttt{mtext()} functions.

Unfortunately, these core tools for drawing text are quite limited
in terms of the formatting of the text.
For example, there is no facility for emphasizing an individual word using
a \textbf{bold} or \emph{italic} face within a text label.

The \CRANpkg{gridtext} \citep{pkg-gridtext} and \CRANpkg{ggtext} \citep{pkg-ggtext}
packages greatly improved the situation by allowing text labels
to include a small subset of
markdown and HTML (plus CSS). This allowed, for example, changes in font face
and color within text labels.

More recently, the \CRANpkg{marquee} package \citep{pkg-marquee}
improved the situation a great deal further by providing full support for
markdown within text labels. This made it possible to layout more
complex arrangements of text and even graphical content within text labels.

\begin{figure}
\includegraphics[width=1\linewidth]{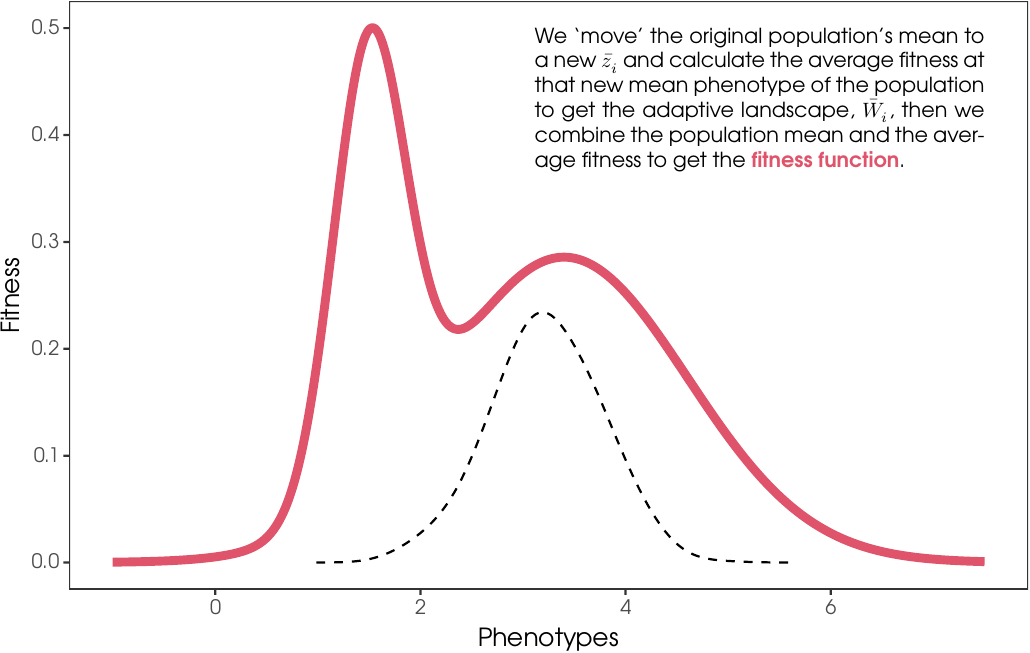} \caption{A plot with a text annotation that contains several typesetting challenges:  in-line mathematical equations; changes in color; and automated line-breaks with full justification and hyphenation.}\label{fig:typesetting}
\end{figure}

However, despite these advances, there are still some text formatting tasks that
remain out of reach. For example, Figure \ref{fig:typesetting}
shows a plot with a text annotation in the top-right corner that contains
a combination of features that cannot be produced using the
available text-drawing tools.

The annotation in Figure \ref{fig:typesetting}
may not appear to be particularly special nor particularly complicated
at first glance, but it harbors several important details:

\begin{itemize}
\item
  The text is a mixture of plain text and mathematical expressions (like
  \(\bar{z}_i\)). Furthermore, the mathematical expressions use a different font
  (Latin Modern) than the plain text (TeX Gyre Adventor)
  and the mixture is broken across multiple lines.

  The R graphics system can draw mathematical expressions \citep{plotmath}
  and that includes a mixture of plain text and mathematical expressions.
  Furthermore, the R graphics system uses a separate symbol font
  for mathematical expressions compared to plain text. However,
  further changes in font within the plain text is not possible and line
  breaks are not supported.
  There is also the problem
  that the typesetting of mathematical expressions in R graphics is not
  of a very high quality.
\item
  The text is not all the same color; the final two words (but not the full
  stop) are red. Furthermore, the final two words are \textbf{bold};
  they have a different
  font face compared to the rest of the text.

  The R graphics system can only draw a character value with a single color
  and a single font face.
  The \CRANpkg{gridtext} and
  \CRANpkg{ggtext} packages make it possible to change color within
  a character value, but they do not allow a mixture of plain text and
  mathematical expressions.
\item
  The text is broken over multiple lines. Furthermore, the text is fully
  justified (not ragged-left or ragged-right justified) and one word
  has been split across lines and hyphenated. Although it is not
  obvious from the plot itself, the line breaks were also automatically
  generated to fit the text into a fixed width.

  The R graphics system can draw a character value across multiple lines, but
  only if explicit newlines are embedded in the character value (i.e., the line
  breaks are manual). The \CRANpkg{gridtext} and
  \CRANpkg{ggtext} packages can calculate simple
  automated line breaks, but they will not break a word across lines (or
  hyphenate) and they cannot fully justify the resulting text. The
  \CRANpkg{marquee} package can automate line breaks and fully justify
  text, but it cannot hyphenate nor can it produce mathematical equations.
\end{itemize}

The features outlined above are all examples of \emph{typesetting}; determining an
arrangement of individual characters and symbols (glyphs) that could be as
simple as placing one character after another (from left to right), but could
also be as complex as arranging mathematical symbols, splitting text into
multiple columns, or writing text vertically from top to bottom.

From R 4.3.0, it has been
possible to draw text from a set of typeset glyphs using the
functions \texttt{grDevices::glyphInfo()} and \texttt{grid::grid.glyph()}
\citep{murrell-pedersen-urbanek-glyphs-2023}. This facility
offers the promise of being able to render arbitrary typeset text
in R. However, it
presupposes that we are able to generate a set of typeset glyphs.

The \CRANpkg{marquee} package provides an example of a package
that can generate typeset glyphs. It is capable of converting
markdown input into a set of glyphs and their positions, which are
then rendered in R.

This article describes the \texttt{xdvir} package, which is another example
of a package that can generate typeset glyphs.
In this case, the input is \LaTeX{}, a \TeX{} engine is used to
generate a set of glyphs and their positions, and then the
result is rendered in R.
The benefit of the \texttt{xdvir} package is that it provides access to the
typesetting capabilities of \LaTeX{}, which includes hyphenation,
fully justified text, mixtures of plain text and mathematical equations---all
of the features demonstrated in Figure \ref{fig:typesetting}---and much more.

The next section describes the basic usage of the \texttt{xdvir} package.
This is followed by a section that breaks down the
design of the \texttt{xdvir} package to show the steps that are
required to render \LaTeX{} output in R.
Subsequent sections explore several of the complexities
that can arise with rendering \LaTeX{} text in R graphics and
some of the solutions that
the \texttt{xdvir} package provides.
The article ends with several extended examples of rendering \LaTeX{} text in R.

\section{\texorpdfstring{\LaTeX{} text labels in R}{ text labels in R}}\label{userinterface}

The simplest way to draw \LaTeX{} text with the \texttt{xdvir} package
is to call the \texttt{grid.latex()} function. The first argument to
this function is a character value, which is interpreted as
a fragment of \LaTeX{} code.
For example, the following code draws a text label that contains
a subset of the larger annotation from
Figure \ref{fig:typesetting}. We use just a subset here in order
to keep the code readable.

Because \LaTeX{} code tends to contain
a large number of backslashes, the code below uses the \texttt{r"(...)"}
syntax for raw character constants, so that we do not have to
escape each backslash with a double backslash.
The resulting image is shown below the code.
Although it is not immediately obvious from that image,
the text, or rather the glyphs, in the image are rendered by R.

\begin{verbatim}
library(xdvir)
\end{verbatim}

\begin{verbatim}
simpleTeX <- r"(We move the original mean to $\bar z_i$)"
\end{verbatim}

\begin{verbatim}
grid.latex(simpleTeX)
\end{verbatim}

\includegraphics[width=0.4\linewidth]{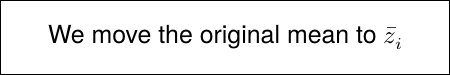}

It is possible to produce something similar to this result using
the \emph{plotmath} feature in R, as shown in the following code (and the image
below the code). However, this demonstrates that
one advantage of using \texttt{xdvir}, even for a simple
piece of text like this, is the superior quality of the \LaTeX{}
fonts and typesetting for mathematical expressions.

\begin{verbatim}
plotmath <- expression("We move the original mean to "*bar(italic(z))[i])
\end{verbatim}

\begin{verbatim}
grid.text(plotmath)
\end{verbatim}

\includegraphics[width=0.4\linewidth]{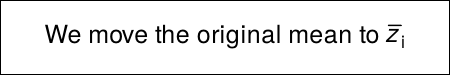}

Another immediate benefit of \texttt{xdvir} is that we can automatically fit
the text within a specified width.
For example, the following code draws the \LaTeX{} fragment \texttt{tex}
again, but this time forces it to fit within a column that is half
the width of the image.

\begin{verbatim}
grid.latex(simpleTeX, width=.5)
\end{verbatim}

\includegraphics[width=0.4\linewidth]{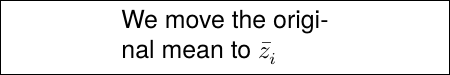}

As the function name \texttt{grid.latex()} suggests, that function
produces low-level
drawing in the \texttt{grid} package graphics system.
The text is just drawn relative to the current \texttt{grid} viewport,
wherever that may be on the page.
While this is extremely flexible, it is more likely that
we want to combine and coordinate the text with a high-level plot
of some sort, like the annotation in Figure \ref{fig:typesetting}.
There are various ways that low-level \texttt{grid} drawing can be combined
with a high-level plot, but we will leave those demonstrations to
later sections.

Instead, for now, we will demonstrate a more common scenario:
drawing \LaTeX{} text labels on a \CRANpkg{ggplot2} plot \citep{pkg-ggplot2}.
For this purpose, the \texttt{xdvir} package provides the
\texttt{element\_latex()} function. This allows us to specify a \LaTeX{} fragment
as a plot label and then we can indicate the special nature of the label
via the \texttt{ggplot2::theme()} function.

\begin{figure}
\includegraphics[width=1\linewidth]{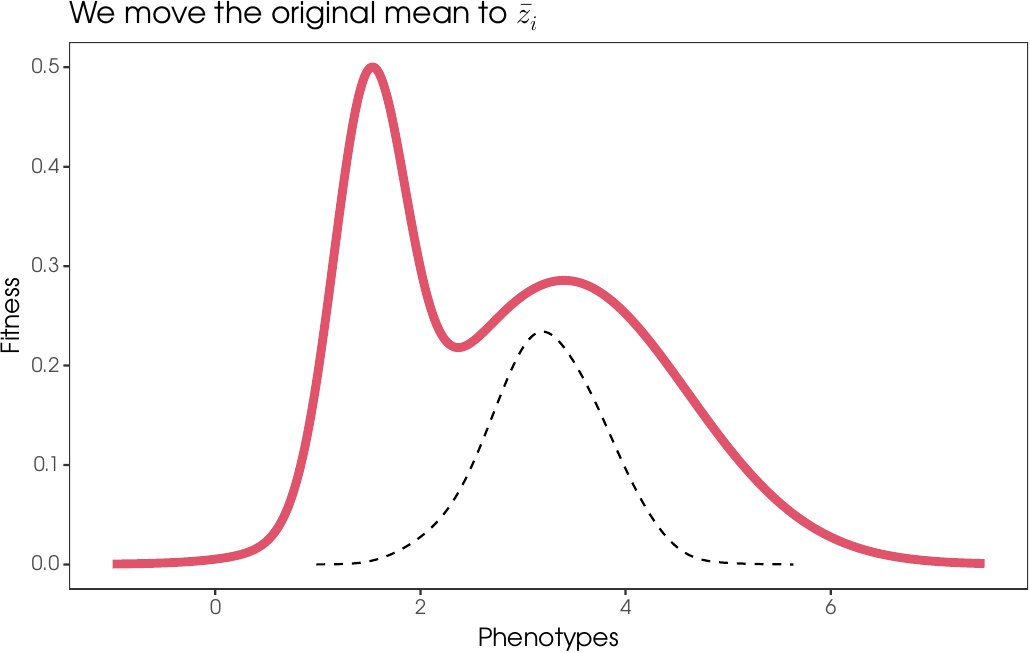} \caption{The \texttt{ggplot2} plot from Figure \ref{fig:typesetting}, without the text annotation, but with a title that was specified using a \LaTeX{} fragment and the function \texttt{element\_latex()}.}\label{fig:elementlatex}
\end{figure}

For example, the following code uses the same \LaTeX{} fragment
from the example above as the title of a \CRANpkg{ggplot2} plot.
The resulting plot is shown in Figure \ref{fig:elementlatex}.
One detail about this result is that the text in this title is larger
than the text drawn by the call to \texttt{grid.latex()} above,
even though exactly the same \TeX{} fragment is being drawn.
A closer inspection reveals that the font is also different.
These differences reflect the fact that \texttt{grid.latex()} and \texttt{element\_latex()}
respect the graphical parameter settings---font families and font sizes---that
are in effect when the \LaTeX{} fragment is drawn. In Figure
\ref{fig:elementlatex} that means respecting the theme settings
of the \CRANpkg{ggplot2} plot.
The \texttt{ggIntro} object in the code below
contains a description of the main \CRANpkg{ggplot2}
plot from Figure
\ref{fig:typesetting}.
The code for generating
\texttt{ggIntro} is not shown in order to keep the code below readable,
but it is available in the supplementary materials for this article.

\begin{verbatim}
library(ggplot2)
\end{verbatim}

\begin{verbatim}
ggIntro +
    labs(title=simpleTeX) +
    theme(plot.title=element_latex())
\end{verbatim}

\begin{figure}
\includegraphics[width=1\linewidth]{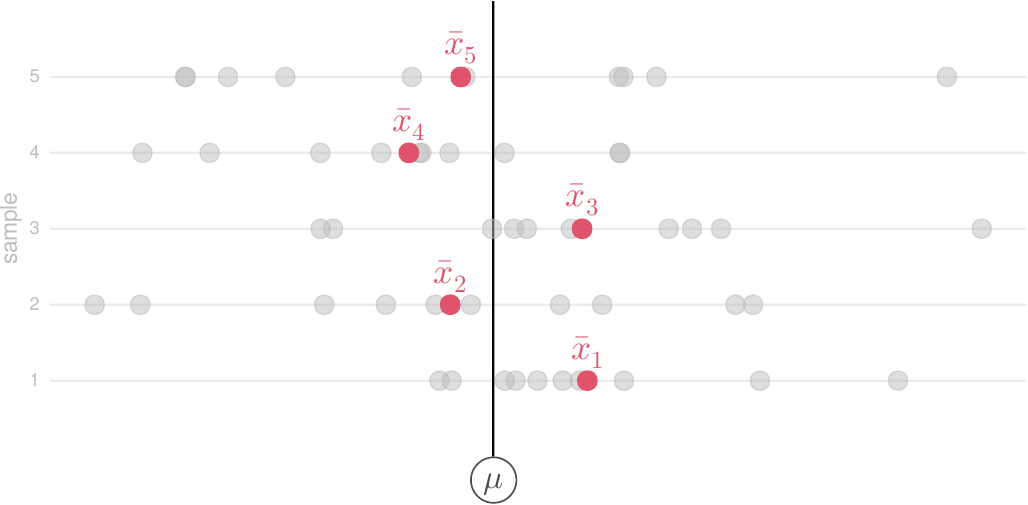} \caption{A \texttt{ggplot2} plot with text labels specified as \LaTeX{} fragments and drawn using the \texttt{geom\_latex()} function.}\label{fig:geomlatex}
\end{figure}

The \texttt{xdvir} package also provides a \texttt{geom\_latex()} function for
drawing text labels, similar to the standard \texttt{ggplot2::geom\_text()} function.
The values specified for the
\texttt{label} aesthetic for \texttt{geom\_latex()} are treated as fragments of \LaTeX{}
code.
For example, Figure \ref{fig:geomlatex}
shows a plot with a set of red points
and a set of red labels, one for each point.
The points are drawn using the standard \texttt{ggplot::geom\_point()} function,
but the labels are drawn using \texttt{geom\_latex()} from the \texttt{xdvir} package.
The red labels for the red points in Figure \ref{fig:geomlatex}
are small \LaTeX{} fragments that each describe
a simple \LaTeX{} mathematical expression.
The data set used for the red points and labels is stored
in a data frame called \texttt{means} and the \LaTeX{} fragments are
in a column called \texttt{label}, as shown below.

\begin{verbatim}
means$label
\end{verbatim}

\begin{verbatim}
#> [1] "$\\bar x_1$" "$\\bar x_2$" "$\\bar x_3$" "$\\bar x_4$" "$\\bar x_5$"
\end{verbatim}

The following code draws the plot in Figure \ref{fig:geomlatex}.
A call to \texttt{ggplot2::geom\_point()} draws the red points and a call
to \texttt{geom\_latex()} draws the red labels.
The \texttt{ggGeom} object in the code below describes the
main plot, which consists of gray dots, horizontal and vertical lines,
and y-axis labels.
The code for generating
\texttt{ggGeom} is not shown in order to keep the code below readable,
but it is available in the supplementary materials for this article.

\begin{verbatim}
ggGeom +
    geom_point(aes(x, sample), data=means, colour=2, size=4) +
    geom_latex(aes(x, sample, label=label), data=means, 
               size=6, vjust=-.4, colour=2)    
\end{verbatim}

\section{Under the hood}\label{underhood}

The previous section showed that simple usage of the \texttt{xdvir} package
only requires specifying a \LaTeX{} fragment as the text to draw.
For example, several examples used the \LaTeX{} fragment shown below.

\begin{verbatim}
simpleTeX
\end{verbatim}

\begin{verbatim}
#> [1] "We move the original mean to $\\bar z_i$"
\end{verbatim}

The \texttt{grid.latex()} function has three tasks to perform in order to
draw that \LaTeX{} fragment in R:

\begin{description}
\item[\textbf{Authoring:}]
The \LaTeX{} fragment has to be turned into a complete
\LaTeX{} document.

The \texttt{author()} function in the \texttt{xdvir} package
allows us to perform this step separately.
For example, the following code
takes the \LaTeX{} fragment \texttt{simpleTeX} and produces a complete \LaTeX{}
document, \texttt{simpleDoc}, that is ready to typeset.
\end{description}

\begin{verbatim}
simpleDoc <- author(simpleTeX)
\end{verbatim}

\begin{verbatim}
simpleDoc
\end{verbatim}

\begin{verbatim}
#> %% R package xdvir_0.1.2; engine name: XeTeX; engine version: XeTeX 3.14159265-2.6-0
#> \documentclass[varwidth]{standalone}
#> \usepackage{unicode-math}
#> \begin{document}
#> We move the original mean to $\bar z_i$
#> \end{document}
\end{verbatim}

\begin{description}
\item[\textbf{Typesetting:}]
The \LaTeX{} document has to be typeset to produce a set of
glyphs and their positions.

The \texttt{typeset()} function in the \texttt{xdvir} package
allows us to perform this step separately.
For example, the following code takes the \LaTeX{} document \texttt{simpleDoc} and
produces a \texttt{"DVI"} object, \texttt{simpleDVI},
that contains instructions specifying the fonts to
use (lines that contain \texttt{x\_fnt\_def} and \texttt{fnt\_num} in the output below),
the glyphs to use from those fonts (lines that contain
\texttt{x\_glyph} in the output below), and where to draw those glyphs
(lines that contain \texttt{down} and \texttt{right} and \texttt{x\_glyph}).
The output shown below has been trimmed to save space and to
make it easier to read.
\end{description}

\begin{verbatim}
simpleDVI <- typeset(simpleDoc)
simpleDVI
\end{verbatim}

\begin{verbatim}
#> pre          version=7, num=25400000, den=473628672, mag=1000,
#>              comment=R package xdvir_0.1.2; engine name: XeTeX; engine version: XeTe
#> bop          counters=1 0 0 0 0 0 0 0 0 0, p=-1
#> xxx1         k=47
#>              x=pdf:pagesize width 143.26802pt height 9.48027pt
#> down3        a=-4114988
#> 
#> ...
#> 
#> right3       b=-4736287
#> x_fnt_def    fontnum=24, ptsize=655360
#>              fontname=/usr/share/texmf/fonts/opentype/public/lm/lmroman10-regular.ot
#> fnt_num_24
#> x_glyph      id=113, x=0, y=0
#> x_glyph      id=50, x=619315, y=0 
#> w3           b=218235 
#> x_glyph      id=75, x=0, y=0
#> 
#> ...
\end{verbatim}

\begin{description}
\item[\textbf{Rendering:}]
The result of the typesetting step has to be drawn in R.

The \texttt{render()} function in the \texttt{xdvir} package
allows us to perform this step separately.
For example, the code below renders the typesetting information from
the \texttt{simpleDVI} object
in R. The resulting image is shown below the code.
\end{description}

\begin{verbatim}
render(simpleDVI)
\end{verbatim}

\includegraphics[width=0.4\linewidth]{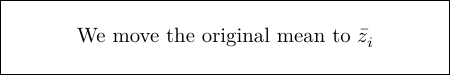}

One detail about the output above is that the rendered text from this
\texttt{render()} call is
smaller and in a different font compared to the example from
the previous section,
which was produced by a \texttt{grid.latex()} call.
This reflects the fact that \texttt{grid.latex()},
in the authoring step, respects the font family and
font size that are in effect when the text is rendered.
By contrast, the \texttt{render()} call is drawing typeset information
from a \LaTeX{} document that just makes use of the default \LaTeX{} font,
Computer Modern (or to be more precise, a modernized version called
Latin Modern) at 10pt.

\section{\texorpdfstring{\LaTeX{} packages}{ packages}}\label{packages}

The code examples so far have dealt with relatively simple fragments of
\LaTeX{} code that consist of just text plus some simple mathematical expressions.
While this is already useful, it barely scratches the surface of
what is possible with
\LaTeX{} code.

Many additional effects can be obtained with \LaTeX{} code by loading
\LaTeX{} packages. As a simple example, changing the color of text
requires loading the \LaTeX{} package \texttt{xcolor}.
These \LaTeX{} packages can be loaded using the \texttt{packages} argument of the
\texttt{grid.latex()} function (or the \texttt{element\_latex()} function
or the \texttt{geom\_latex()} function).
For example, the following code draws text with the last two
words in red.

\begin{verbatim}
colourTeX <- r"(We combine to get the \color{red}{Fitness Function})"
\end{verbatim}

\begin{verbatim}
grid.latex(colourTeX, packages="xcolor")
\end{verbatim}

\includegraphics[width=0.4\linewidth]{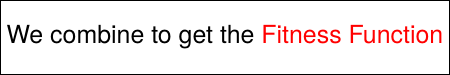}

The argument \texttt{packages="xcolor"} is used in the authoring step to load the
package in the \LaTeX{} document preamble. This is demonstrated below
with an explicit call to the \texttt{author()} function. We can see that
\texttt{\textbackslash{}usepackage\{xcolor\}} has been added to the \LaTeX{} document.

\begin{verbatim}
colourDoc <- author(colourTeX, packages="xcolor")
\end{verbatim}

\begin{verbatim}
colourDoc
\end{verbatim}

\begin{verbatim}
#> %% R package xdvir_0.1.2; engine name: XeTeX; engine version: XeTeX 3.14159265-2.6-0
#> \documentclass[varwidth]{standalone}
#> \usepackage{unicode-math}
#> \usepackage{xcolor}
#> \begin{document}
#> We combine to get the \color{red}{Fitness Function}
#> \end{document}
\end{verbatim}

This in turn affects the typesetting step: without the \texttt{xcolor} package, the
\LaTeX{} command \texttt{\textbackslash{}color} would not be recognized; with the \texttt{xcolor} package,
the \texttt{\textbackslash{}color} command produces instructions to change color
in the \texttt{"DVI"} output. This is demonstrated below with an
explicit call to the \texttt{typeset()} function. An example of the color-change
instructions is the line containing \texttt{color\ push} in the output below the code.

\begin{verbatim}
colourDVI <- typeset(colourDoc)
colourDVI
\end{verbatim}

\begin{verbatim}
#> pre          version=7, num=25400000, den=473628672, mag=1000,
#>              comment=R package xdvir_0.1.2; engine name: XeTeX; engine version: XeTe
#> bop          counters=1 0 0 0 0 0 0 0 0 0, p=-1
#> 
#> ...
#> 
#> x_fnt_def    fontnum=24, ptsize=655360
#>              fontname=/usr/share/texmf/fonts/opentype/public/lm/lmroman10-regular.ot
#> fnt_num_24
#> x_glyph      id=113, x=0, y=0
#> x_glyph      id=50, x=619315, y=0 
#> 
#> ...
#> 
#> xxx1         k=20
#>              x=color push rgb 1 0 0
#> x_glyph      id=54, x=0, y=0
#> x_glyph      id=66, x=427950, y=0
#> 
#> ...
\end{verbatim}

The argument \texttt{packages="xcolor"} is also
used in the rendering step because, without it,
the rendering would not take any notice of the instructions to
change color. This is demonstrated below with an explicit
call to the \texttt{render()} function. The resulting image differs from
the previous one because it uses the default \LaTeX{} font, but we can see the
same change in color for the last two words.

\begin{verbatim}
render(colourDVI, packages="xcolor")
\end{verbatim}

\includegraphics[width=0.4\linewidth]{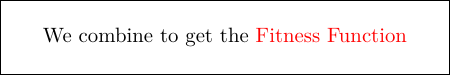}

There are several \LaTeX{} packages with predefined support in the
\texttt{xdvir} package, including
\texttt{xcolor} for changes in color and \texttt{fontspec} for changes in font.
Support can be added for other \LaTeX{} packages with the
\texttt{LaTeXpackage()} function. We will see other predefined packages
and an example of defining a new \LaTeX{}
package in later sections.

\section{Justifying text}\label{justifying-text}

By default, the \LaTeX{} text drawn by \texttt{grid.latex()}
is centered upon a specified location.
For example, the following code draws the \texttt{simpleTeX}
fragment vertically centered
at a location half-way up the image (as indicated by the gray line).

\begin{verbatim}
grid.latex(simpleTeX, y=.5)
\end{verbatim}

\includegraphics[width=0.4\linewidth]{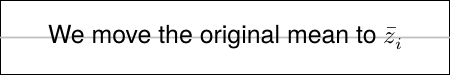}

We can specify a different justification using the \texttt{vjust} argument.
For example, the following code draws the same \texttt{simpleTeX} fragment
at the same location, but with a bottom-justification.
Notice that the bottom of the text is based on the bounding box
of the text, so the bottom of the text is the bottom of the subscript ``i''.

\begin{verbatim}
grid.latex(simpleTeX, y=.5, vjust="bottom")
\end{verbatim}

\includegraphics[width=0.4\linewidth]{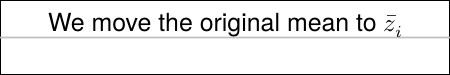}

In some situations it will be much more useful to
justify text relative to the text baseline, as shown by the following code.

\begin{verbatim}
grid.latex(simpleTeX, y=.5, vjust="baseline")
\end{verbatim}

\includegraphics[width=0.4\linewidth]{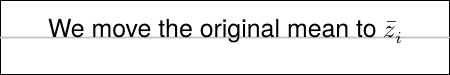}

The \texttt{xdvir} package has a very simple algorithm for determining the text
baseline, but there is also predefined support for the \LaTeX{} package \texttt{preview},
which produces a more reliable baseline. That baseline can be accessed,
assuming the \texttt{preview} package is loaded, by
specifying \texttt{vjust="preview-baseline"}.

There is also an \texttt{hjust} argument for horizontal justification.
This accepts the standard values, \texttt{"left"}, \texttt{"centre"}, and \texttt{"right"}, but
also accepts \texttt{"bbleft"}, \texttt{"bbcentre"}, and \texttt{"bbright"}.
The latter three are based on a bounding box around the actual ink
that is drawn, which does not include space before or after glyphs
(left-side bearing and right-side bearing).
The following code provides a demonstration of the difference
by drawing the simple \LaTeX{} fragment from previous examples as the title
of a ggplot2 plot.
We add a (mathematical) vertical bar to the end of the \LaTeX{} fragment
and draw the title larger than normal and justify the text against the
right side of the plot region, using \texttt{"right"} justification first and
then using \texttt{"bbright"} justification.
The output below the code just shows the very top of the plot in order to save
space.

\begin{verbatim}
rightBearingTeX <- paste0(simpleTeX, "$|$")
\end{verbatim}

\begin{verbatim}
ggIntro + 
    labs(title=rightBearingTeX) +
    theme(plot.title=element_latex(size=20, hjust="right"))
\end{verbatim}

\includegraphics[width=0.95\linewidth]{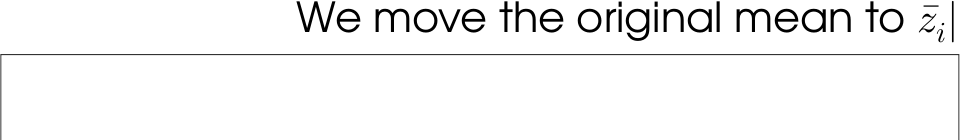}

\begin{verbatim}
ggIntro + 
    labs(title=rightBearingTeX) +
    theme(plot.title=element_latex(size=20, hjust="bbright"))
\end{verbatim}

\includegraphics[width=0.95\linewidth]{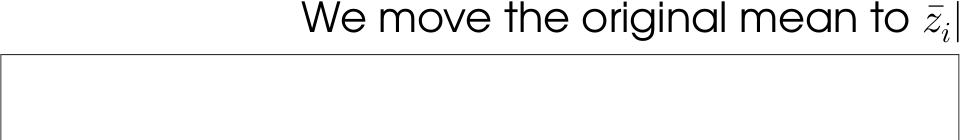}

The difference between the two plots is that the second vertical bar is
precisely aligned with the right edge of the plot region whereas the
first vertical bar is slightly to the left of the right edge of the plot
region (because of the right-side bearing of the vertical bar glyph).
This is a very small detail, but it is something that can be visually
jarring if we are trying to align components of a plot in order
to produce a clean design. This fine level of control is exactly
the sort of precision that we are seeking by working with \LaTeX{} typesetting.

\section{Integrating text}\label{integrating-text}

Justifying \LaTeX{} text is a simple example of a larger problem:
\emph{integrating} \LaTeX{} text. For example,
the text annotation in Figure \ref{fig:typesetting}
is integrated with the plot in the sense that it is positioned
relative to the plot region. In fact, closer inspection reveals
that the text annotation
is carefully top-justified with the maximum y-value of the
red line and right-justified with the maximum x-value of the red line.

Put in terms of \emph{integration} rather than justification,
the text annotation in Figure \ref{fig:typesetting}
is integrated with the plot because the \LaTeX{} text is drawn at a location
that is coordinated with the location of other R graphics drawing in the plot.

Another example of integration, that reverses the roles, is coordinating
other R graphics drawing with the location of \LaTeX{} text.
The following code provides a simple example.
The \LaTeX{} fragment is the simple one from previous examples with
two additions: there are \texttt{\textbackslash{}zsavepos} commands to mark specific locations
within the text and associate them with labels (\texttt{"a"} and \texttt{"b"});
and there are \texttt{\textbackslash{}Rzmark} commands to export those
locations for R to see.

\begin{verbatim}
zrefTeX <- r"(We move the original\zsavepos{a} mean to \zsavepos{b}$\bar z_i$
\Rzmark{a}\Rzmark{b})"
\end{verbatim}

If we render this \LaTeX{} fragment, we just get the familiar output.
The commands that we added to the \LaTeX{} fragment
are based on the \LaTeX{} package \texttt{zref},
so we must load that package.

\begin{verbatim}
grid.latex(zrefTeX, packages="zref")
\end{verbatim}

\includegraphics[width=0.4\linewidth]{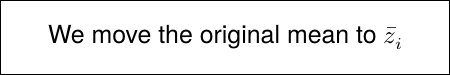}

However, we can now access the special locations in the \LaTeX{}
output using the
\texttt{getMark()} function from the \texttt{xdvir} package.
For example, the following code accesses location \texttt{"a"}, which is just after
the word ``original'', and draws
a small red dot at that location.

\begin{verbatim}
a <- getMark("a")
grid.circle(a$devx, a$devy, r=unit(.5, "mm"), gp=gpar(col=2, fill=2))
\end{verbatim}

\includegraphics[width=0.4\linewidth]{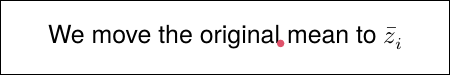}

The following code accesses location \texttt{"b"}, which is just before the letter ``z'',
and draws a curved arrow
from \texttt{"a"} to \texttt{"b"}.

\begin{verbatim}
b <- getMark("b")
grid.xspline(unit.c(a$devx, .5*(a$devx + b$devx), b$devx),
             unit.c(a$devy, a$devy - unit(3, "mm"), a$devy),
             shape=-1, gp=gpar(col=2, fill=2),
             arrow=arrow(length=unit(2, "mm"), type="closed"))
\end{verbatim}

\includegraphics[width=0.4\linewidth]{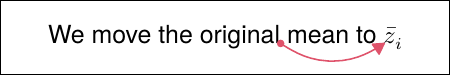}

The exported locations also produce ``anchors'' that we can use to justify
\LaTeX{} text. For example, the following code
draws the simple \LaTeX{} fragment with position \texttt{"a"} at
the center of the image (which is indicated by gray lines).

\begin{verbatim}
grid.latex(zrefTeX, packages="zref", hjust="a", vjust="a")
\end{verbatim}

\includegraphics[width=0.4\linewidth]{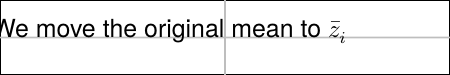}

\begin{figure}
\includegraphics[width=1\linewidth]{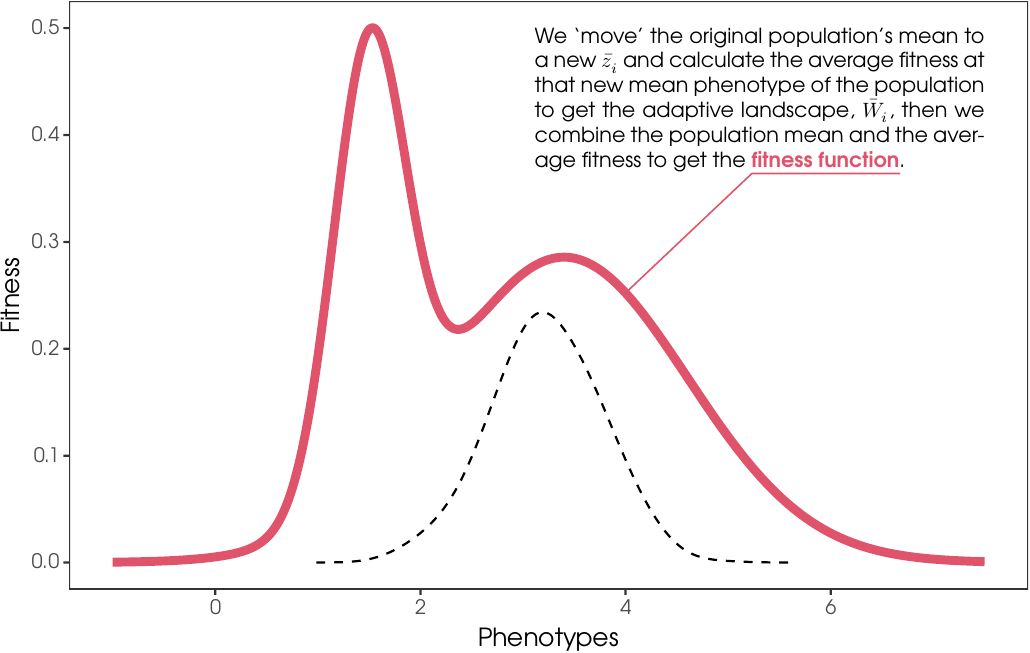} \caption{The \texttt{ggplot2} plot from Figure \ref{fig:typesetting}, including the \LaTeX{} annotation, with a line added relative to marked locations within the \LaTeX{} annotation (and relative to the thick red line).}\label{fig:zref}
\end{figure}

Figure \ref{fig:zref} provides a more realistic demonstration.
This figure shows the plot from Figure \ref{fig:typesetting}
with a line added to visually connect the thick red line with
the red part of the \LaTeX{} annotation.
The code for this plot is not shown for reasons of space, but it
makes use of the same basic idea as the code above by saving locations within
the \LaTeX{} output and then accessing them with the \texttt{getMark()} function.
The full code is available in the supplementary materials for this article.

\section{\texorpdfstring{\LaTeX{} graphics}{ graphics}}\label{graphics}

The examples so far have demonstrated using \LaTeX{} code to describe
text labels, combined with using R to draw general
graphics---lines and circles and so on.
It is also possible to use \LaTeX{} to draw general graphics.
In particular, the \LaTeX{} package Ti\textit{k}Z provides very powerful and flexible
graphics facilities.
The \texttt{xdvir} package provides support for the \LaTeX{} package Ti\textit{k}Z,
so we are able to render Ti\textit{k}Z graphics in R.

For example, the following \LaTeX{} code describes a Ti\textit{k}Z picture
consisting of two text labels enclosed within circles, with
arrows connecting the circles.

\begin{verbatim}
tikzTeX <- r"(%
\path (0, 0) node[circle,minimum size=.5in,draw,thick] (x) {\sffamily{R}} 
       (3, 0) node[circle,minimum size=.5in,draw,thick] (y) {Ti{\it k}Z!};
\draw[-{stealth},thick] (x) .. controls (1, 1) and (2, 1).. (y);
\draw[-{stealth},thick] (y) .. controls (2, -1) and (1, -1) .. (x);)"
\end{verbatim}

The following code draws this Ti\textit{k}Z picture in R.
The argument
\texttt{packages="tikzPicture"} is necessary to ensure that the Ti\textit{k}Z package
is loaded in the authoring step, that Ti\textit{k}Z output is produced in the
typesetting step, and that R takes notice of the Ti\textit{k}Z
output in the rendering step.

\begin{verbatim}
grid.latex(tikzTeX, packages="tikzPicture")
\end{verbatim}

\includegraphics[width=0.4\linewidth]{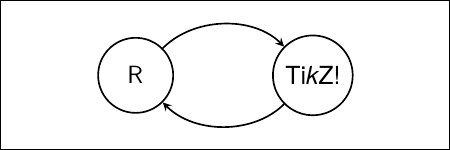}

The label on the x-axis of Figure \ref{fig:geomlatex}
is another simple Ti\textit{k}Z picture that
uses Ti\textit{k}Z commands to draw the Greek letter mu
within a circle.
This example is not completely trivial because it uses the
\LaTeX{} concept of ``phantom'' text to make the circle large enough to fit
a capital ``M'' even though no such character is drawn.
This is another example of the detailed typsetting capabilities
that access to \LaTeX{} provides.

\begin{verbatim}
muDot <- r"(%
\begin{tikzpicture}
\node[draw,circle,thick,inner sep=0.5mm]{\vphantom{M}$\mu$};
\end{tikzpicture})"
\end{verbatim}

The \LaTeX{} code this time includes an explicit \texttt{\textbackslash{}begin\{tikzpicture\}}
and \texttt{\textbackslash{}end\{tikzpicture\}}. Those commands were implicitly added in the previous
example because we specified \texttt{packages="tikzPicture"}.
This time, we have explicitly provided the commands, so we
just specify \texttt{packages="tikz"}.

\begin{verbatim}
grid.latex(muDot, packages="tikz")
\end{verbatim}

\includegraphics[width=0.4\linewidth]{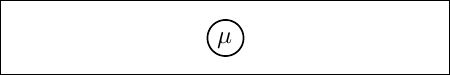}

We will see a more complex example of Ti\textit{k}Z output in a later section.
Figure \ref{fig:diag} is also a Ti\textit{k}Z picture that has been rendered
in R.

\section{\texorpdfstring{Programmatic generation of \LaTeX{}}{Programmatic generation of }}\label{programmatic-generation-of}

Although \LaTeX{} fragments
for text labels tend to be more complex than
plain text labels, thanks to the additional markup that is required,
\LaTeX{} code is still
just text. This means that all of the text-generating tools in R
are available to help with authoring \LaTeX{} fragments.
For example, the labels used to render text data symbols in
Figure 3 could be generated via a simple call to the \texttt{paste0()} function,
as shown below.

\begin{verbatim}
paste0("$\\bar x_", 1:5, "$")
\end{verbatim}

\begin{verbatim}
#> [1] "$\\bar x_1$" "$\\bar x_2$" "$\\bar x_3$" "$\\bar x_4$" "$\\bar x_5$"
\end{verbatim}

There are also packages that can generate larger fragments of
\LaTeX{} code. For example, there are packages like \CRANpkg{xtable}
\citep{pkg:xtable}
for generating \LaTeX{} tables and the \CRANpkg{rmarkdown}
package \citep{pkg:rmarkdown}
can generate \LaTeX{} documents from Markdown input.
These tools can be useful for generating larger chunks of \LaTeX{} code,
although the \LaTeX{} code that is produced may consist of
entire documents rather than just \LaTeX{} fragments.
The next section describes how we can cope with that situation.

\section{Customization and debugging}\label{customization-and-debugging}

\begin{figure}
\includegraphics[width=1\linewidth]{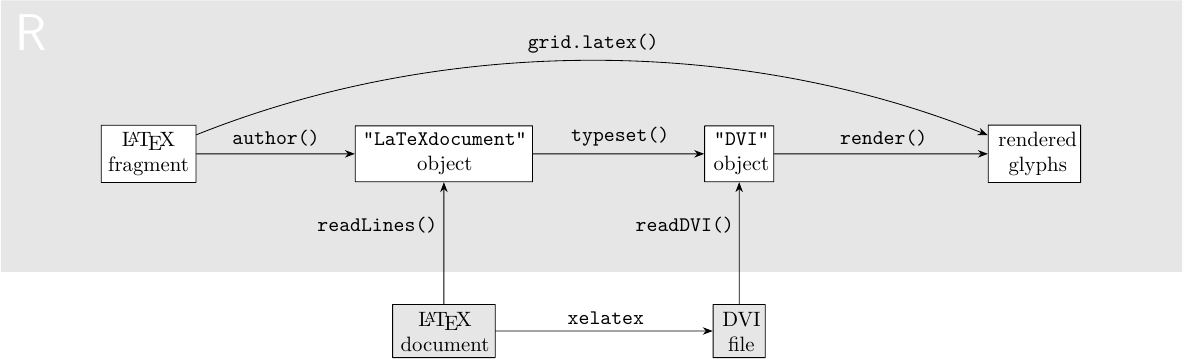} \caption{The design of the `xdvir` package.}\label{fig:diag}
\end{figure}

Most of the examples in this article take a fragment of \LaTeX{} code
and pass it to the \texttt{grid.latex()} function, which performs an
authoring step, a typesetting step, and a rendering step.
We saw in a previous section that there are functions
\texttt{author()}, \texttt{typeset()}, and \texttt{render()}
that allow us to perform these steps separately (see Figure \ref{fig:diag}).
This provides more control over the individual steps and allows
us to inspect the results of the individual steps, which can be
useful for debugging.
In this section, we explore further options for controlling the
authoring, typesetting, and rendering steps.

The \texttt{author()} function transforms a \LaTeX{} fragment into a
complete \LaTeX{} document. Although there are arguments to the \texttt{author()}
function that allow some control over that transformation,
e.g., the \texttt{packages} argument, it does not allow full control
over the composition of the \LaTeX{} document.
Fortunately, a \LaTeX{} document within R is essentially just a character vector,
so another way to author a \LaTeX{} document is to create an external text file
and read that into R. This allows complete control over the content
of the \LaTeX{} document.
Another possibility is that we want to use a \LaTeX{} document that
we did not create, for example, if we write Markdown code and
convert it to \LaTeX{} code.

The \texttt{typeset()} function transforms a \LaTeX{} document into a \texttt{"DVI"} object
that contains a set of typeset glyphs. There is limited control over this
process as well, with only the \texttt{engine} argument allowing us to select between
\texttt{"xetex"} or \texttt{"luatex"}. Again, one way to
obtain greater control is to
perform this step outside of R by running
a \TeX{} engine, e.g., \texttt{xelatex}, on an external text file to produce a DVI
file. The \texttt{xdvir} package provides the \texttt{readDVI()} function
to read external DVI files into R and these can then be passed to
the \texttt{render()} function for drawing.

One important caveat is that both
a \texttt{"LaTeXdocument"} object that is produced by
the \texttt{author()} function and a \texttt{"DVI"} object that is produced by the
\texttt{typeset()} function contain information about how they were created,
for example, the \TeX{} \texttt{engine} that was specified and the \LaTeX{}
packages that were loaded.
The \texttt{typeset()} function checks this information and warns if we ask
to typeset a \texttt{"LaTeXdocument"} that was produced for a different \TeX{} engine.
Similarly, the \texttt{render()} function, which also has an \texttt{engine} argument,
checks and warns if we ask to render a
\texttt{"DVI"} object that was produced using a different \TeX{} engine.

External \LaTeX{} documents and DVI files do not (explicitly) contain this
information so it is up to the user to ensure that the \TeX{} engine,
and any \LaTeX{} packages, are consistent with the arguments provided
to the functions \texttt{typeset()} and \texttt{render()}.
In some situations, even with the appropriate level of care,
it will be impossible to avoid warnings.

\section{Example 1}\label{example-1}

\begin{figure}
\includegraphics[width=1\linewidth]{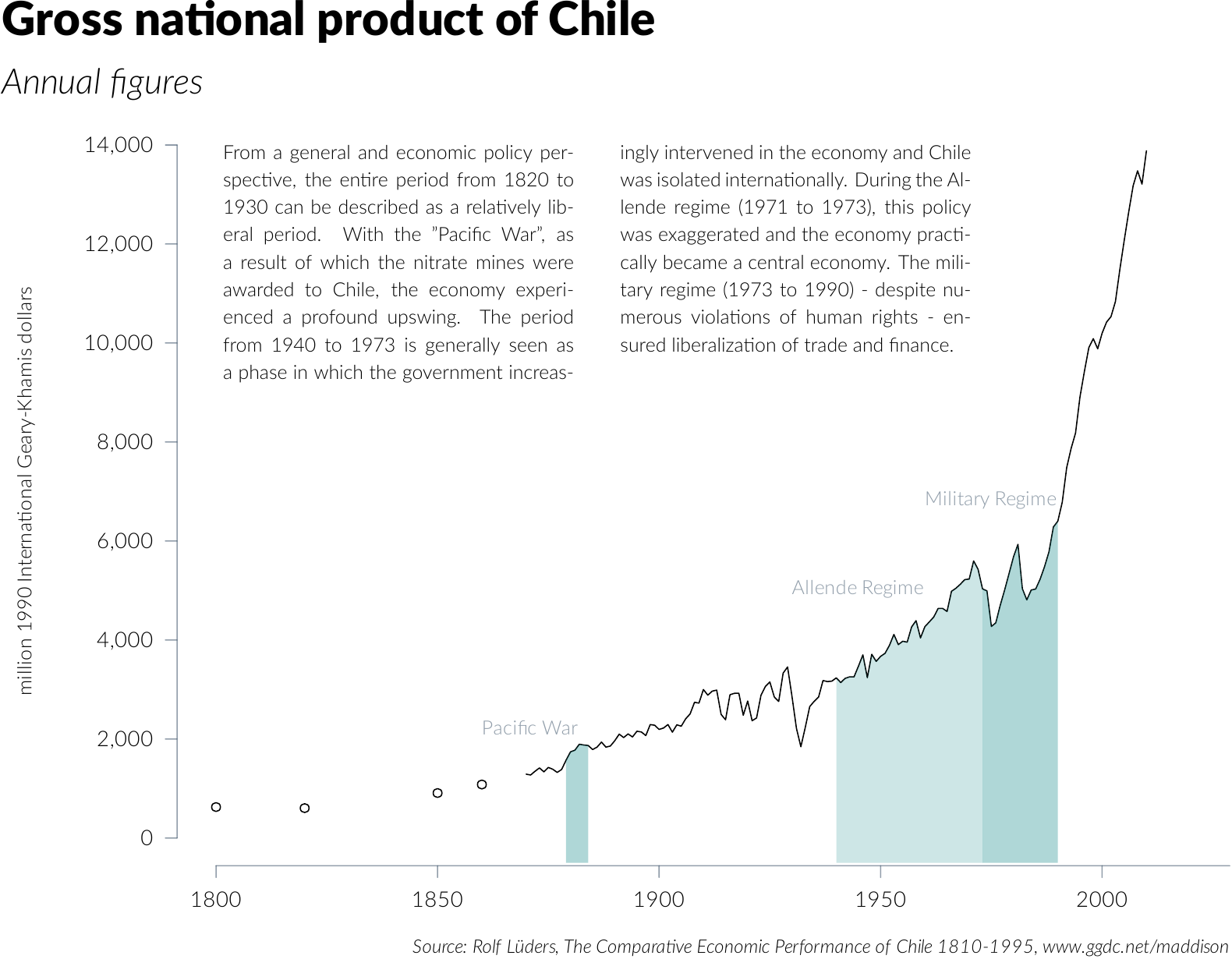} \caption{A plot with a two-column text annotation. This plot is an adaptation of Figure 4.1 from Thomas Rahlf's book ``Data Visualisation with R'' \citep{rahlf}.}\label{fig:rahlf}
\end{figure}

This section demonstrates a more complete example of rendering \LaTeX{} text
within a plot. The plot, shown in Figure \ref{fig:rahlf},
provides a clear example of the more advanced typesetting
capabilities of \LaTeX{}; the text annotation in the top-left corner
of the plot is not only typeset into two columns,
but both columns are fully justified and feature several examples of
hyphenation.

This example also demonstrates one way to integrate a
\texttt{grid.latex()} call with a plot that was drawn using functions
from the \texttt{graphics} package. We will also see a simple
demonstration of the \texttt{LaTeXpackage()} function to allow use of
a \LaTeX{} package that has no predefined support in \texttt{xdvir}.

\begin{figure}
\includegraphics[width=1\linewidth]{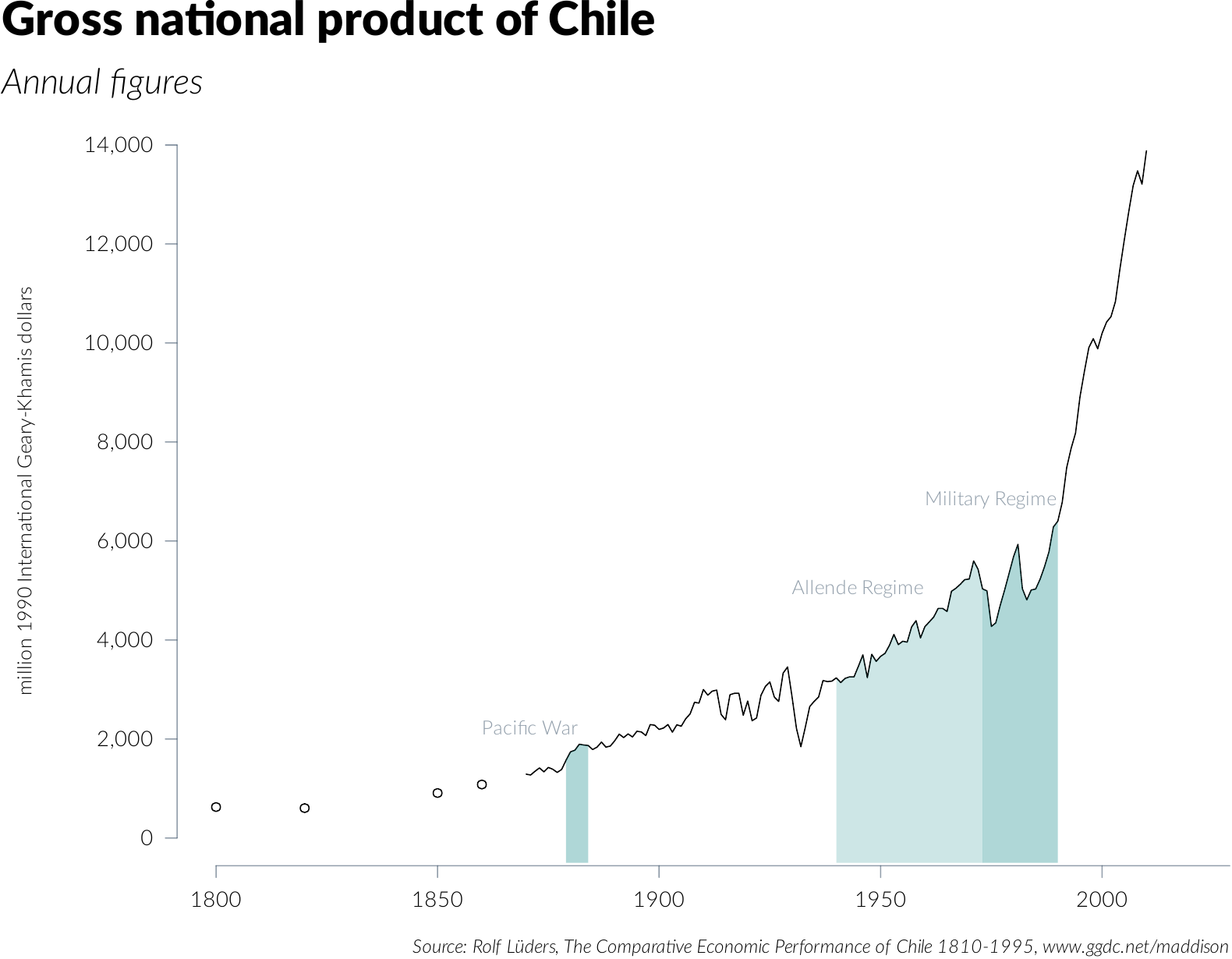} \caption{The main plot from Figure \ref{fig:rahlf} without the two columns of text annotation. This plot is drawn using functions from the \texttt{graphics} package.}\label{fig:rahlfplain}
\end{figure}

The details of the code that produces the main plot---everything except
the two columns of text in the top-left corner---are not relevant
to this article so we perform this drawing
just with a call to a \texttt{rahlfPlot()} function that is defined in the
supplementary material for the article. The result is shown in
Figure \ref{fig:rahlfplain}.

\begin{verbatim}
rahlfPlot()
\end{verbatim}

Because the main plot is drawn using functions from the \texttt{graphics} package,
in order to integrate the output from \texttt{grid.latex()} with the plot,
we need to convert the plot to an equivalent drawing that uses
functions from the \texttt{grid} package.
This can be achieved with the \texttt{grid.echo()} function from the
\CRANpkg{gridGraphics} package \citep{pkg-gridgraphics}, as shown below.

\begin{verbatim}
library(gridGraphics)
grid.echo()
\end{verbatim}

We want to integrate
the \LaTeX{} text with the main plot. In particular, we want
the top of the text to be aligned with the value 14,000 on the y-scale
of the plot. There is also a 1cm gap between the left of the text
and the y-axis line.
In order to achieve this, we can navigate to the \texttt{grid} viewport
that corresponds to the main plot region, which also has scales
that match the plot scales.
The naming scheme for the \texttt{grid} viewports that \texttt{grid.echo()}
generates is described in \citet{RJ-2015-012}.

\begin{verbatim}
downViewport("graphics-window-1-1")
\end{verbatim}

We are now ready to render the \LaTeX{} text within the plot.
The \LaTeX{} code for this example is shown below.
This is a larger \LaTeX{} fragment than we have previously seen,
but more importantly it contains a larger number of \LaTeX{} commands
to control the typesetting of the text.
For example, we control the font family with a \texttt{\textbackslash{}setmainfont} command,
we control font size and vertical line spacing
with a \texttt{\textbackslash{}fontsize} command,
we control the overall width of the text using a \texttt{minipage}
environment,
we set the number of columns using a \texttt{multicol} environment,
and we control the horizontal spacing between columns with
a \texttt{\textbackslash{}setlength} command.

\begin{verbatim}
#> \setmainfont{Lato-Light}
#> \fontsize{12pt}{17pt}\selectfont
#> \setlength{\columnsep}{1cm}
#> \begin{minipage}[t]{16.25cm}
#> \begin{multicols}{2} 
#> From a general and economic policy perspective, the entire period from
#> 1820 to 1930 can be described as a relatively liberal period. With the
#> "Pacific War", as a result of which the nitrate mines were awarded to
#> Chile, the economy experienced a profound upswing. The period from
#> 1940 to 1973 is generally seen as a phase in which the government
#> increasingly intervened in the economy and Chile was isolated
#> internationally. During the Allende regime (1971 to 1973), this policy
#> was exaggerated and the economy practically became a central
#> economy. The military regime (1973 to 1990) - despite numerous
#> violations of human rights - ensured liberalization of trade and
#> finance.
#> \end{multicols}
#> \end{minipage}
\end{verbatim}

The \texttt{\textbackslash{}setmainfont} and
\texttt{\textbackslash{}fontsize} commands in the \LaTeX{} code require the \LaTeX{} package \texttt{fontspec}
to be loaded, but this is not a problem because there is predefined support
for \texttt{fontspec} in the \texttt{xdvir} package. However, the \texttt{multicol} environment
in the \LaTeX{} code requires the \LaTeX{} package \texttt{multicol} and there is no
predefined support for that in \texttt{xdvir}. The following code uses the
\texttt{LaTeXpackage()} function to provide support for the \LaTeX{} package \texttt{multicol}.
In a simple case like this, all we have to do is provide a name for the
package (\texttt{"multicol"}) and use the \texttt{preamble} argument to
provide the \LaTeX{} code that should be added in the
authoring step to load the \LaTeX{} package.
We also call the \texttt{registerPackage()} function so that we can refer to this \LaTeX{}
package by its name.

\begin{verbatim}
multicol <- LaTeXpackage("multicol",
                         preamble="\\usepackage{multicol}")
registerPackage(multicol)
\end{verbatim}

Finally, we call \texttt{grid.latex()} to add the \LaTeX{} text to the plot.
The object \texttt{rahlfTeX} contains the \LaTeX{} code,
we specify the \LaTeX{} packages that have to be loaded, including
the \texttt{"multicol"} package that we just registered, and
we position the text 1cm in from the left of the
the plot viewport and at 14,000 on the y-axis.
The final result is shown in Figure \ref{fig:rahlf}.

\begin{verbatim}
grid.latex(rahlfTeX, 
           packages=c("fontspec", "multicol"),
           x=unit(1, "cm"), y=unit(14000, "native"), 
           hjust="left", vjust="top")
\end{verbatim}

\section{Example 2}\label{example-2}

\begin{figure}
\includegraphics[width=1\linewidth]{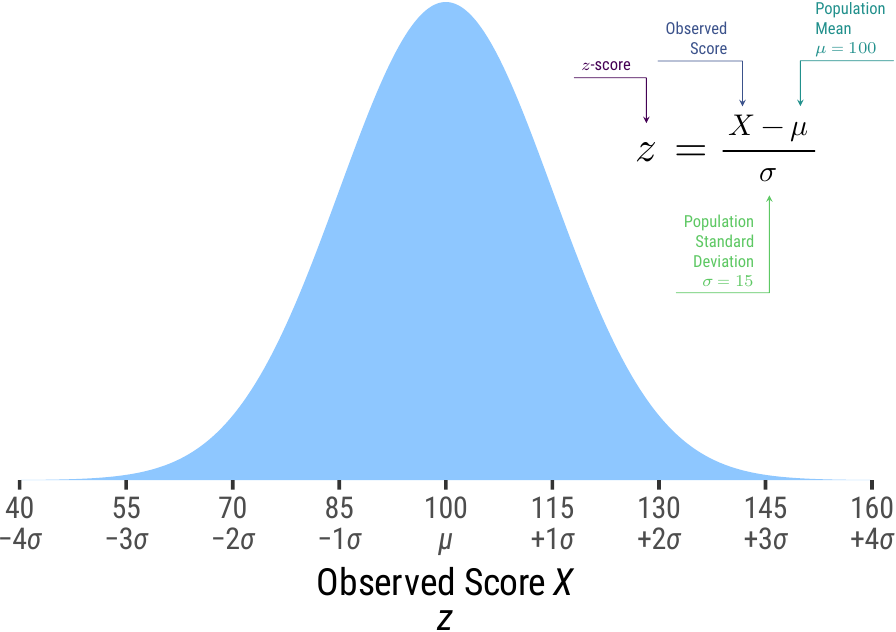} \caption{A plot with annotated mathematical equation. This plot is an adaptation of the plot in \citet{schneider2023}.}\label{fig:schneider}
\end{figure}

This section looks at another more complete example of a plot with
a \LaTeX{} annotation (Figure \ref{fig:schneider}). This example
demonstrates the sophisticated effects that are possible by
combining Ti\textit{k}Z graphics with \LaTeX{} typesetting, in this case
to produce an annotated mathematical equation.
This example also demonstrates a way to integrate lower-level
\texttt{grid.latex()} output with a \CRANpkg{ggplot2} plot
(rather than using \texttt{element\_latex()} or \texttt{geom\_latex()}).

\begin{figure}
\includegraphics[width=1\linewidth]{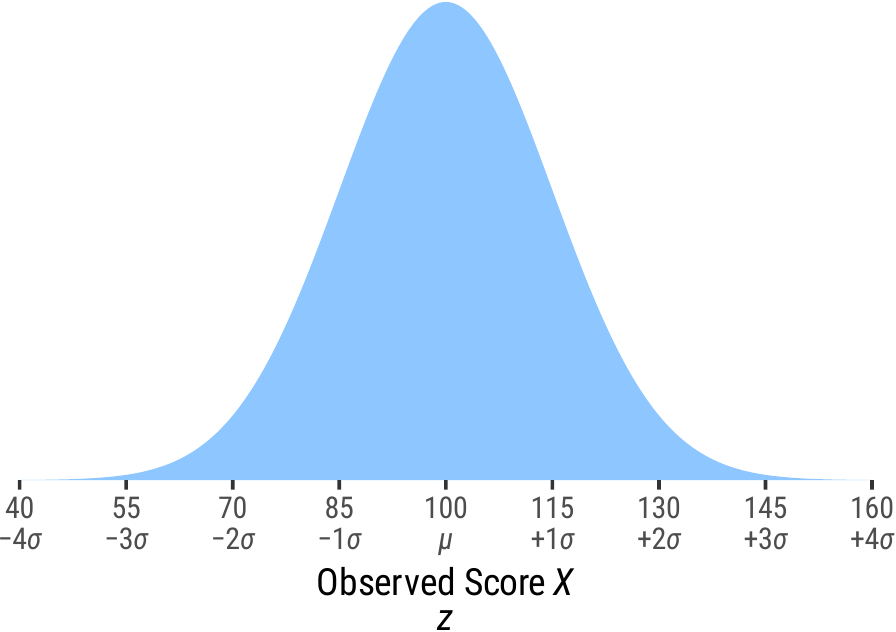} \caption{The main plot from Figure \ref{fig:schneider} without the annotated mathematical equation. This plot is produced using the packages \CRANpkg{ggplot2} and \CRANpkg{ggtext}.}\label{fig:schneiderplain}
\end{figure}

The main plot in this example is a \CRANpkg{ggplot2} plot.
The details of the code that generates the main plot
are not particularly relevant to this article, so the
main plot is described in the object \texttt{ggSchneider}, which
is defined in the supplementary materials for the article.
One point worth noting is that the labeling on the x-axis, which
combines italic Greek letters with upright digits and signs, is produced
using the \CRANpkg{ggtext} package.
In other words, this example combines two levels of text annotation:
labels on the x-axis that are relatively simple, but still beyond the
capabilities of core R text drawing; and much more sophisticated
text annotations that require access to a complex system like \LaTeX{}.
The main plot produced by \texttt{ggSchneider} is shown in Figure
\ref{fig:schneiderplain}.

\begin{verbatim}
ggSchneider
\end{verbatim}

The start of the \LaTeX{} code for the annotated equation is shown below
(the full code is included in the supplementary materials for this article).
The \LaTeX{} code
is arranged in three blocks: the first block of code defines some colors;
the second block describes the main mathematical equation, but includes
some \texttt{\textbackslash{}eqnmark} commands to save locations within the equation;
and the third block shows one of the additional equation annotations, which
refers to one of the saved locations within the main mathematical
equation, in this case
the ``z'', and positions a
label relative to that location, in this case the label ``z-score'',
which is positioned
above and to the left of the ``z''.

\begin{verbatim}
#> \definecolor{myviolet}{HTML}{440154}
#> \definecolor{myblue}{HTML}{3B528B}
#> \definecolor{myindigo}{HTML}{21908C}
#> \definecolor{mygreen}{HTML}{5DC863}
#>     
#> \huge$
#> \eqnmark[myviolet]{z}{z} = 
#> \frac{
#>     \eqnmark[myblue]{x}{X}-
#>     \eqnmark[myindigo]{mu}{\mu}}{
#>     \eqnmark[mygreen]{sigma}{\sigma}}
#> $
#> 
#> \annotate[
#>     yshift=1em, 
#>     myviolet,
#>     align=right]
#>     {above, left}
#>     {z}
#>     {$z$-score}
#> 
\end{verbatim}

There are several \LaTeX{} packages required by this \LaTeX{} code,
in particular the \texttt{\textbackslash{}eqnmark} and \texttt{\textbackslash{}annotate} commands require the \LaTeX{}
package \texttt{annotate-equations}.
As in the previous example, we can add support for this package
using the \texttt{LaTeXpackage()} and \texttt{registerPackage()} functions.
One difference this time is that the \texttt{annotate-equations} package is being
loaded from a local \texttt{TeX} directory.

\begin{verbatim}
annotateEquations <-
    LaTeXpackage(name="annotate",
                 preamble="\\usepackage{TeX/annotate-equations}")
registerPackage(annotateEquations)
\end{verbatim}

The \LaTeX{} package \texttt{annotate-equations} is built on Ti\textit{k}Z graphics.
We do not need to load the \LaTeX{} package \texttt{tikz} because \texttt{annotate-equations}
will do that automatically. However, \texttt{xdvir} by default makes use
of the bounding box information from Ti\textit{k}Z graphics and, for images with
saved locations like this, that bounding box is unreliable.
The predefined support for the \LaTeX{} package \texttt{tikz} in the \texttt{xdvir} package
includes a \texttt{tikzPackage()} function that allows us to load TikZ, but
ignore its bounding boxes, as shown in the following code.

\begin{verbatim}
tikzNoBBox <-
    tikzPackage(name="tikzNoBBox", bbox=FALSE)
registerPackage(tikzNoBBox)
\end{verbatim}

Finally, we will use the \LaTeX{} package \texttt{roboto} to access
specific variations of the Roboto font for the text labels
in the annotated equation.

\begin{verbatim}
roboto <-
    LaTeXpackage(name="roboto",
                 preamble="\\usepackage[sfdefault,condensed]{roboto}")
registerPackage(roboto)
\end{verbatim}

Rendering the annotated equation on the plot
requires integrating the \LaTeX{} output with the
\CRANpkg{ggplot2} plot. In particular, we want to align the top of the \LaTeX{}
output with the top of the density curve and we want to align the right side
of the \LaTeX{} output with the right edge of the label ``160'' on the x-axis.

We saw in an earlier section
how to use \texttt{element\_latex()} to draw \LaTeX{} text
in labels such as the plot title on a \CRANpkg{ggplot2} plot
and how to use \texttt{geom\_latex()} to draw \LaTeX{} text
as data symbols.
In this example, we are adding a single \LaTeX{} annotation at a specific
position within a \CRANpkg{ggplot2} plot, so
we use the \CRANpkg{gggrid} package \citep{pkg-gggrid}.
This package provides the \texttt{grid\_panel()} function, which we can
add to a \CRANpkg{ggplot2} plot, much like the standard \texttt{ggplot2::geom\_point()}
function, to add \texttt{grid} drawing to a \texttt{ggplot2} plot.
The first argument to \texttt{grid\_panel()} is a function that must
generate a \texttt{grid} grob for \CRANpkg{ggplot2} to draw, based on the
data values that \CRANpkg{ggplot2} passes to it.
In this case, we define a function called \texttt{annotation()}, which
calls the \texttt{xdvir} function \texttt{latexGrob()}. The \texttt{latexGrob()} function
is similar to
\texttt{grid.latex()} except that it creates a description of something to draw
rather than immediately drawing it.
We pass to \texttt{latexGrob()} the \LaTeX{} code to draw the annotated
equation (\texttt{schneiderTeX}),
a set of \texttt{packages} to load, and arguments that position
the output relative to the plot.
The final result is shown in Figure \ref{fig:schneider}.

\begin{verbatim}
library(gggrid)

annotation <- function(data, coords) {
    latexGrob(schneiderTeX, 
              packages=c("tikzNoBBox", "annotate", "roboto"),
              x=unit(coords$x, "npc") + 0.5*stringWidth("160"),
              y=coords$y, hjust=1, vjust=1)
}

ggSchneider +
    grid_panel(annotation, 
               aes(x=x, y=y), 
               data=data.frame(x=160, y=dnorm(100, mean=100, sd=15)))
\end{verbatim}

\section{Example 3}\label{example-3}

\begin{figure}
\includegraphics[width=1\linewidth]{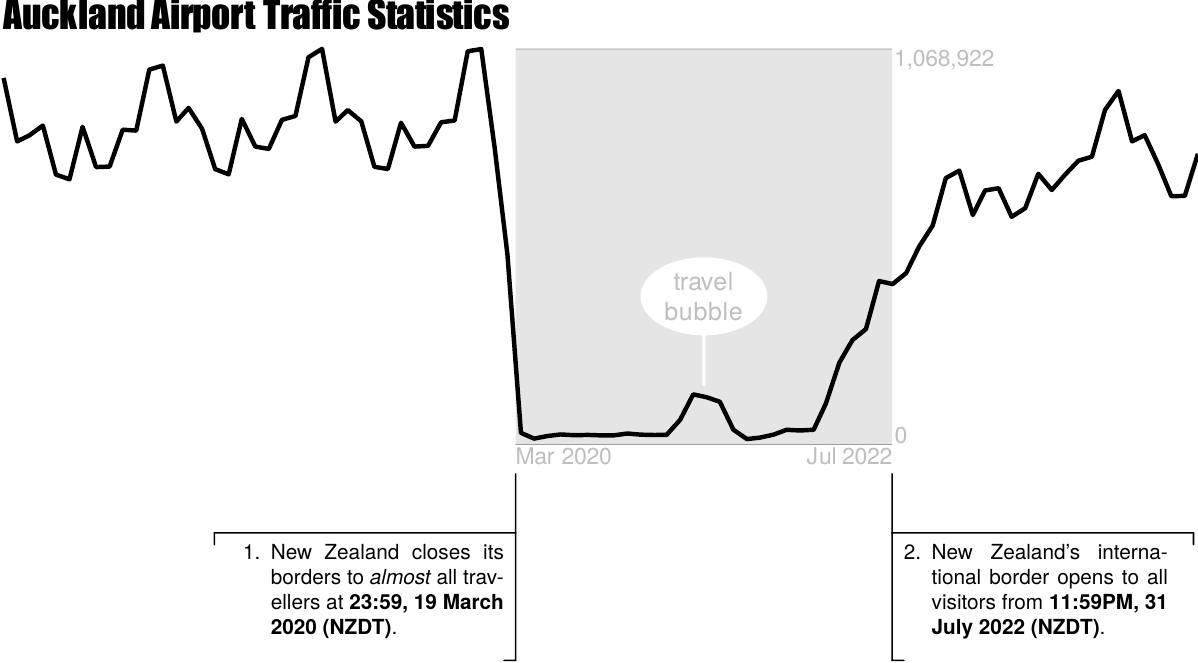} \caption{A plot with numbered list items as annotations below the plot. This plot is an adaptation of Figure 10 from \citet{anzjs2025}.}\label{fig:anzjs}
\end{figure}

This section
provides another demonstration of the range of possibilities that is
provided by
\LaTeX{} typesetting. This
time we add annotations that are formatted as numbered
list items below a plot (Figure \ref{fig:anzjs}).

\begin{figure}
\includegraphics[width=1\linewidth]{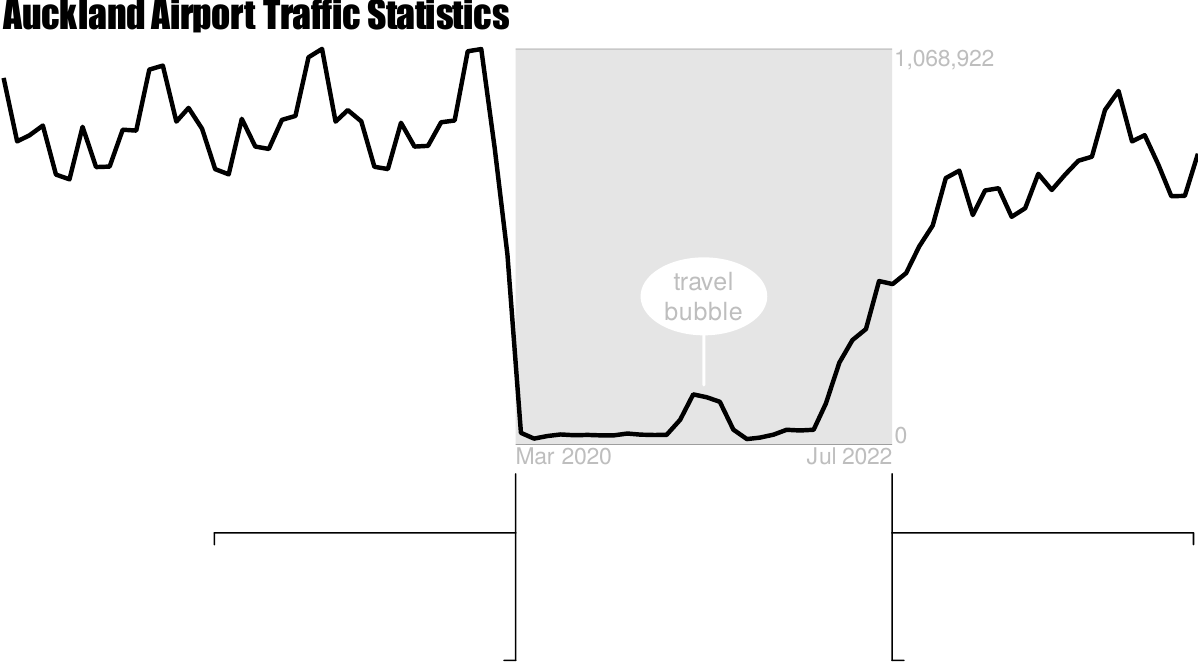} \caption{The main plot from Figure \ref{fig:anzjs} without the numbered list items as annotations. This plot is produced using the \CRANpkg{ggplot2} package.}\label{fig:anzjsplain}
\end{figure}

The main plot is a \CRANpkg{ggplot2} plot with a number of relatively
simple annotations already added.
The details of the code are not particularly relevant to this
article, so the main plot is described in the object \texttt{ggANZJS},
which is defined in the supplementary materials for the article.
One point worth noting is that the \LaTeX{} annotations that we will
be adding are required to fit within the lines that extend below the plot.
In other words, we will be specifying a fixed width for the \LaTeX{} output
to fit into.
The main plot produced by \texttt{ggANZJS} is shown in Figure \ref{fig:anzjsplain}.

\begin{verbatim}
ggANZJS
\end{verbatim}

We will focus on drawing just the left-hand \LaTeX{} annotation.
The \LaTeX{} code is shown below.
This includes commands to control the font size and an \texttt{enumerate}
environment that creates a numbered list item.

\begin{verbatim}
#> %
#> \fontsize{10}{12}
#> \selectfont
#> \begin{enumerate}
#> \item New Zealand closes its borders to {\it almost} all travellers at
#> {\bf 23:59, 19 March 2020 (NZDT)}.
#> \end{enumerate}
\end{verbatim}

As with the previous example, we have a single annotation that we want
to position quite carefully, so we define a function that generates
a \texttt{grid} grob to use with the \texttt{grid\_panel()} function from the
\CRANpkg{gggrid} package.
The \texttt{labelLeft()} function calls \texttt{latexGrob()}, gives it the \LaTeX{} code
to draw (\texttt{closeTeX}), specifies the position for the \LaTeX{} output,
and specifies a \texttt{width} for the output to be typeset within.

\begin{verbatim}
labelLeft <- function(data, coords) {
    x1 <- coords$x[1]
    x2 <- coords$x[2]
    w <- unit(1 - x2, "npc") - unit(1, "mm")
    gap <- 15
    latex1 <- latexGrob(closeTeX,
                        x=unit(x1, "npc") - unit(2, "mm"), 
                        y=unit(0, "npc") - unit(gap, "mm") - unit(2, "mm"),
                        hjust=1, vjust=1,
                        width=w)
}
\end{verbatim}

The following code combines the left-hand label annotation,
and a very similar right-hand label annotation, with the
\texttt{ggANZJS} plot.
The final result is shown in Figure \ref{fig:anzjs}.

\begin{verbatim}
ggANZJS +
    grid_panel(labelLeft,
               aes(x=borders),
               data=data.frame(borders=c(borderClosed, borderOpen))) +
    grid_panel(labelRight,
               aes(x=borders),
               data=data.frame(borders=c(borderClosed, borderOpen)))
\end{verbatim}

\section{Example 4}\label{example-4}

\begin{figure}
\includegraphics[width=1\linewidth]{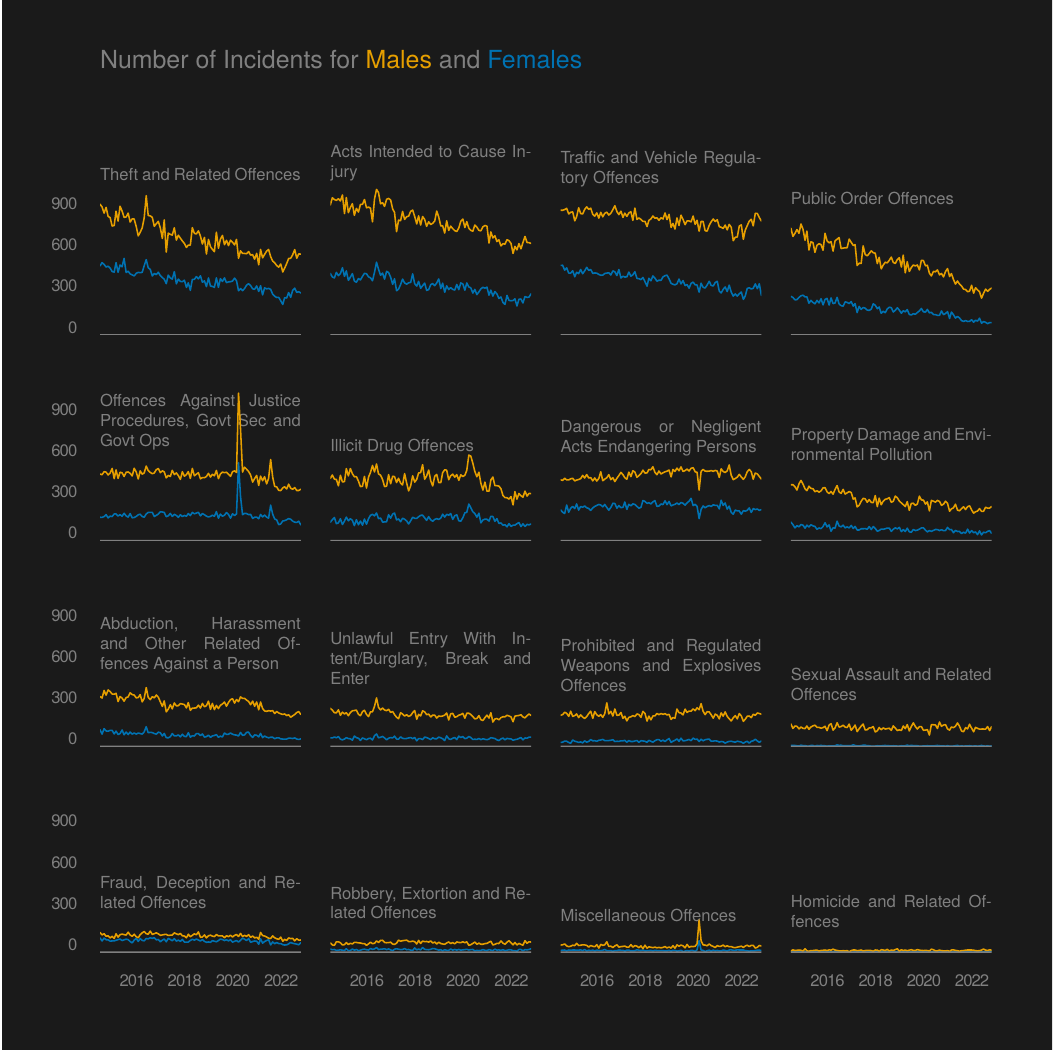} \caption{A \CRANpkg{lattice} plot with \LaTeX{} text used for the plot title and for annotations in each panel.}\label{fig:lattice}
\end{figure}

This section provides an example of integrating \texttt{grid.latex()} output
with a multi-panel \CRANpkg{lattice} plot \citep{pkg-lattice}.
The plot is shown in
Figure \ref{fig:lattice}.

\begin{figure}
\includegraphics[width=1\linewidth]{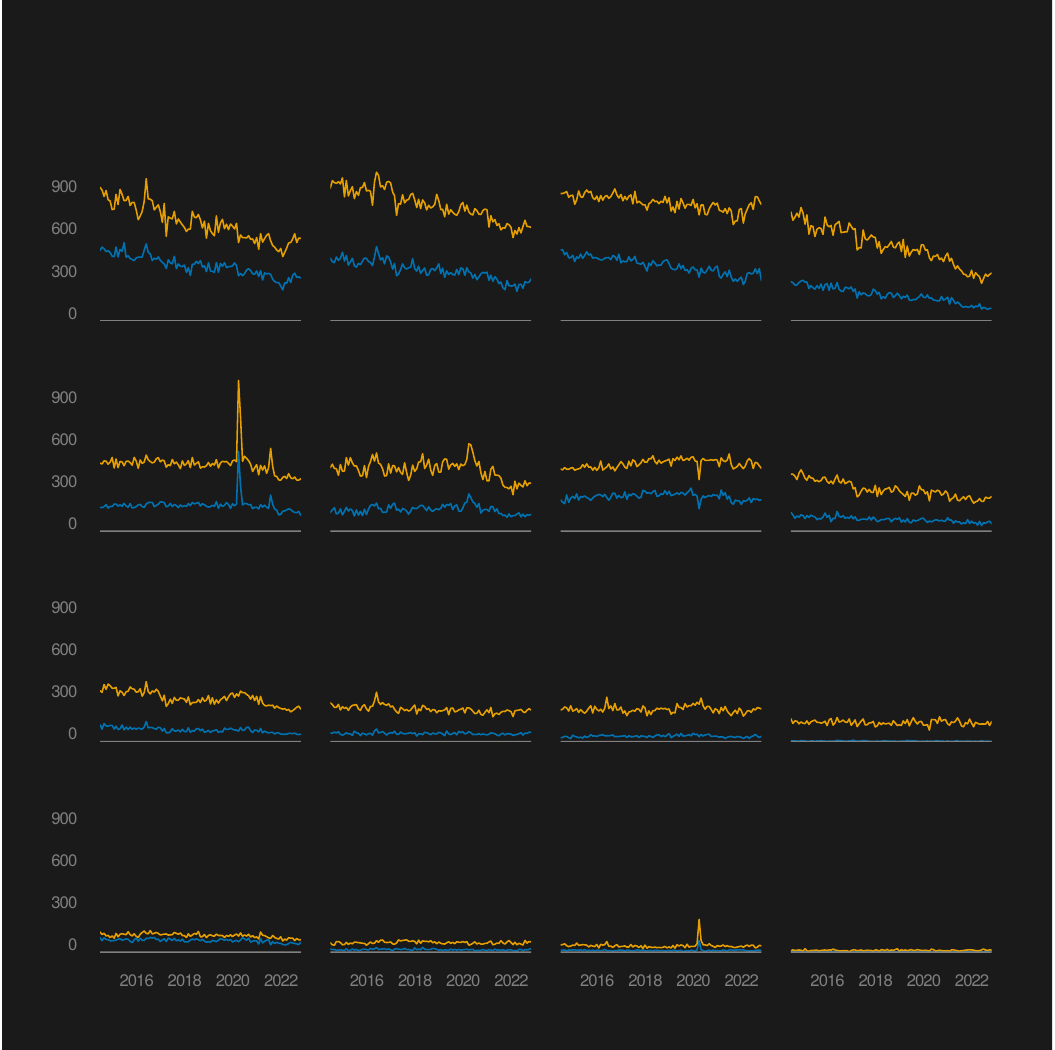} \caption{The main plot from Figure \ref{fig:lattice} without the title and annotations in each panel. This plot is produced using the \CRANpkg{lattice} package.}\label{fig:latticeplain}
\end{figure}

The main plot is a \CRANpkg{lattice} plot consisting of multiple panels,
with separate lines for males and females.
The details of the code for generating the main plot are not relevant
to this article, so it is described in the object \texttt{latticeCrime}, which
is defined in the supplementary material.
The main plot produced by \texttt{latticeCrime} is shown in
Figure \ref{fig:latticeplain}.

\begin{verbatim}
latticeCrime
\end{verbatim}

We can add drawing to each panel of a \CRANpkg{lattice} plot by providing
a \emph{panel function}. The panel function is passed the
relevant data for the panel, and the code within the panel function is run
in the panel viewport, which means that the appropriate axis scales are
available. This means that we can include a call to \texttt{grid.latex()}
within a panel function in order to add \LaTeX{} text to each panel.
For example, the following code defines the panel function for
Figure \ref{fig:lattice}.
This function calculates the appropriate label for the panel
and encloses that within a \LaTeX{} \texttt{minipage} environment
that is the width of the panel. This means that the label is typeset
to be fully-justified within the panel (unless it is a single line that is
narrower than the panel).
We use a \texttt{minipage} environment in the \LaTeX{} fragment
rather than just using the
\texttt{width} argument to \texttt{grid.latex()} because \texttt{minipage} produces a more
precise width.
The panel function then calls \texttt{grid.latex()} to draw
that \LaTeX{} fragment, placing the label slightly above the first data value
for males. The call to the \texttt{mainPanel()} function draws the yellow and
blue lines that are part of the main plot.

\begin{verbatim}
latexPanel <- function(x, y, subscripts, groups, ...) {
    type <- crime$Type[subscripts][1]
    labelY <- y[groups == "Male"][1]
    labelWidth <- convertWidth(unit(1, "npc"), "in", valueOnly=TRUE)
    panelTeX <- paste0("\\begin{minipage}{", labelWidth, "in}",
                       type, 
                       "\\end{minipage}")
    grid.latex(panelTeX, 
               x=0, hjust="left",
               y=unit(labelY, "native") + unit(4, "mm"), vjust="bottom",
               gp=gpar(col=lightGrey, fontsize=8))
    mainPanel(x, y, subscripts, groups, ...)
}
\end{verbatim}

The title of a \CRANpkg{lattice} plot can be specified as a \texttt{grid}
grob. This means that we can call \texttt{latexGrob()} to generate a
title for the plot in Figure \ref{fig:lattice}.
The \LaTeX{} fragment below describes the label, first defining
three colors, and then giving the title text, with the words
``Male'' and ``Female'' colored differently.

\begin{verbatim}
titleTeX <- r"(%
\definecolor{lightGrey}{RGB}{128,128,128}
\definecolor{lattice1}{RGB}{230,159,0}
\definecolor{lattice2}{RGB}{0,114,178}
\color{lightGrey}
Number of Incidents for {\color{lattice1}Males} and {\color{lattice2}Females}
)"
\end{verbatim}

The following code calls \texttt{latexGrob()} to define the title.
We pass the \LaTeX{} fragment \texttt{titleTeX}, we position the title
to line up with
the left edge of the first column of panels, and we
load the \LaTeX{} package \texttt{xcolor} so that the colors work.

\begin{verbatim}
latexTitle <- latexGrob(titleTeX, x=titleX, hjust="left", 
                        packages="xcolor")
\end{verbatim}

The following code creates the final plot by adding the panel function
\texttt{latexPanel} and the title \texttt{latexTitle} to the main plot \texttt{latticeCrime}.
The final result is shown in Figure \ref{fig:lattice}.

\begin{verbatim}
update(latticeCrime,
       panel=latexPanel,
       main=latexTitle)
\end{verbatim}

\section{Discussion}\label{discussion}

The \texttt{xdvir} package provides convenient high-level functions for
rendering \LaTeX{} fragments as labels, annotations, or data symbols on
R plots. The package also provides lower-level functions that
allow more fine control over the authoring, typesetting, and
rendering of \LaTeX{} code in R.

The benefit of the \texttt{xdvir} package is access to the typesetting capabilities
of \LaTeX{}. This ranges from relatively simple features like changes in
font family, font weight, and font style, and automatic line breaks,
to intermediate features like full justification, hyphenation, and high-quality
mathematical equations, and more advanced features like
enumerated lists, multiple columns, and Ti\textit{k}Z graphics.

One limitation of the \texttt{xdvir} package is that rendering \LaTeX{}
fragments is noticeably slower than rendering simple character values.
This is mainly because the typesetting
step requires running a \TeX{} engine to produce a DVI file.
The \texttt{xdvir} package performs some caching in order to minimize the problem,
but the time cost can still be quite large. For example,
Figure \ref{fig:lattice} requires running a \TeX{} engine 17 times.

Another limitation of the \texttt{xdvir} package is that it requires a
graphics device that can render typeset glyphs. This currently includes
the \texttt{pdf()} and \texttt{quartz()} devices, plus all devices based on the
Cairo graphics library \citep{cairo}, and graphics devices
provided by the \texttt{ragg} package \citep{pkg:ragg}.

A final major limitation of \texttt{xdvir} is that it only currently supports two
\TeX{} engines: \XeTeX{} and recent Lua\TeX{}.
The package start-up message
reports on whether these are available. An implicit limitation is that
\texttt{xdvir} requires a \TeX{} installation, though that is simplified
through a dependency on the \CRANpkg{tinytex} package \citep{pkg:tinytex}.

Given these limitations, it is worth discussing alternative approaches.
The first section of this article mentioned \CRANpkg{gridtext},
\CRANpkg{ggtext}, and \CRANpkg{marquee}. These packages
provide alternative ways to render non-trivial text labels, but
do so through Markdown and/or HTML rather than \LaTeX{}.
Although they may not be able to produce as wide a range of results
compared to \LaTeX{} code, they will perform much faster and
require fewer dependencies than \texttt{xdvir}.
There are also a number of packages that perform specific
text-placement tasks, for example \CRANpkg{geomtextpath} \citep{pkg:geomtextpath},
which can arrange text along an arbitrary path, and
\CRANpkg{directlabels} \citep{pkg:directlabels} and
\CRANpkg{ggforce} \citep{pkg:ggforce}, which provide functions for cleverly
positioning text annotations, though without typesetting facilities.
The advantage of \texttt{xdvir} by comparison with these packages
is that it is possible to
produce more advanced typesetting results thanks to having access to \LaTeX{}.

The \CRANpkg{tikzDevice} package \citep{pkg:tikzDevice} is an interesting
alternative because, where \texttt{xdvir} integrates \LaTeX{} text with R graphics,
\CRANpkg{tikzDevice} reverses the process and integrates R graphics with \LaTeX{}.
The \CRANpkg{tikzDevice} package provides an R graphics device that
converts R plots into Ti\textit{k}Z pictures so that R plots can include labels with
\LaTeX{} fragments and R plots can be
deeply integrated with \LaTeX{} documents.
The main difference with this package is the destination:
if we use \texttt{xdvir}, we end up with \LaTeX{} output within an R plot;
if we use \CRANpkg{tikzDevice} we end up with an R plot within \LaTeX{} output.
If the final destination is a \LaTeX{} document,
then the \CRANpkg{tikzDevice} may provide more convenience
and greater control. However, if the final destination is
more general, or unknown, then \texttt{xdvir} may be the more appropriate
solution.

The \CRANpkg{latex2exp} package \citep{pkg:latex2exp} is another package
that works in the opposite direction to \texttt{xdvir}. This package
takes a \LaTeX{} fragment and converts it to an R \emph{plotmath} expression.
This allows users familiar with \LaTeX{} to access R's math-drawing facility
whereas \texttt{xdvir} allows users to access \LaTeX{}'s math-drawing facility,
which is far superior. The advantage of \CRANpkg{latex2exp}, as with
many of these alternative approaches, is that it does not have any
system dependencies, whereas \texttt{xdvir} requires a \TeX{} installation.

Another alternative approach to including \LaTeX{} output in R plots
is to import an image of the \LaTeX{} output.
This approach harks back to early solutions for including
\LaTeX{} mathematical expressions in web pages by generating PNG images
from \LaTeX{} fragments.
However, more modern technologies, such as SVG, mean that this
approach can yield a much higher-quality result, as
demonstrated by \citet{schneider2023}.
One simple advantage of the \texttt{xdvir} approach is the level of convenience
that it provides by automating the authoring and typesetting steps.
The \texttt{xdvir} package also provides more possibilities to integrate
\LaTeX{} output with other drawing in R through anchors and saved positions.

Some of the limitations of \texttt{xdvir} may also be overcome by further
development. For example, it may be possible to extend support
to more \TeX{} engines and to more graphics devices. Providing support
for more \LaTeX{} packages is another area for
future work.

\section{Acknowledgments}\label{acknowledgments}

The \texttt{xdvir} package depends on Yihui Xie's \CRANpkg{tinytex} package
for the typesetting step. This package makes it much
simpler to make use of \TeX{} engines,
including performing multiple runs when necessary, and much easier
to install
\LaTeX{} packages (and \TeX{} itself).

Claus O. Wilke's \CRANpkg{ggtext} package and Thomas Lin Pedersen's
\CRANpkg{marquee} package provided excellent templates for the integration
of improved text-drawing facilities with \CRANpkg{ggplot2}.

The author owes a debt of gratitude to Marc-Olivier Beausoleil
(Figure \ref{fig:typesetting}),
Thomas Rahlf (Figure \ref{fig:rahlf}),
and Joel Schneider (Figure \ref{fig:schneider})
for sharing their work and
for giving either implicit or explicit
permission to base several of the examples in this article
on their work.

\bibliography{murrell-xdvir.bib}

\address{%
Paul Murrell\\
The University of Auckland\\%
Department of Statistics\\ Auckland, New Zealand\\
\url{https://www.stat.auckland.ac.nz/~paul/}\\%
\textit{ORCiD: \href{https://orcid.org/0000-0002-3224-8858}{0000-0002-3224-8858}}\\%
\href{mailto:paul@stat.auckland.ac.nz}{\nolinkurl{paul@stat.auckland.ac.nz}}%
}

\end{article}

\end{document}